%% file: conference_101719.tex
\documentclass[10pt,conference]{IEEEtran}
\AtBeginDocument{%
  \providecommand\BibTeX{{%
    \normalfont B\kern-0.5em{\scshape i\kern-0.25em b}\kern-0.8em\TeX}}}

\usepackage[linesnumbered,ruled,vlined]{algorithm2e}
\usepackage{amsmath,amssymb,amsfonts}
\usepackage{textcomp}
\usepackage{xcolor}
\usepackage{caption}
\usepackage{subcaption}
\usepackage{multirow}
\usepackage{graphicx}
\usepackage{booktabs}
\usepackage{makecell}
\usepackage{enumitem}
\usepackage{listings}
\usepackage{breakcites}
\usepackage{tcolorbox}
\usepackage{multicol} 
\usepackage{subcaption}
\usepackage{ulem}
\usepackage{wrapfig}
\usepackage[noend]{algpseudocode}
\usepackage{amsmath}
\usepackage{cite}
\usepackage{hyperref}

\begin{document}

\include{tex/macros}
%\usepackage{algpseudocode} 
%\usepackage[dvipsnames]{xcolor}

% \renewcommand\theFancyVerbLine{\small\arabic{FancyVerbLine}}
% \setminted{fontsize=\footnotesize,
%     xleftmargin=8mm,
%     linenos,
%     frame=lines,
%     breaklines}

%\settopmatter{printacmref=false}

%\documentclass[sigconf,review,anonymous]{acmart}
%\documentclass[10pt,conference]{IEEEtran}

% \newcommand{\todo}[1]{\textbf{\color{red}{Ravishka: #1} }}

% \newcommand{\wei}[1]{\textcolor{blue}{Wei: [#1]}}

% \newcommand{\ww}[1]{\textcolor{blue}{Weihang: [#1]}}

% \newcommand{\td}[1]{\textcolor{blue}{ToDo: [#1]}}

% \newcommand{\zijie}[1]{{\color{purple}[Zijie: #1]}}

% \newcommand{\zeqing}[1]{{\textcolor{blue}[Zeqing: #1]}}

% \newcommand{\jiajun}[1]{{\color{cyan}[Jiang: #1]}}

% \author{
%     \IEEEauthorblockN{Ravishka Rathnasuriya}
%     \IEEEauthorblockA{University of Texas at Dallas, United States, ravishka.rathnasuriya@utdallas.edu}
   
% }

\author{
\IEEEauthorblockN{Ravishka Rathnasuriya}
\IEEEauthorblockA{ravishka.rathnasuriya@utdallas.edu\\
The University of Texas at Dallas}
}

\title{A Framework for On the Fly Input Refinement for Deep Learning Models}
%\title{CodeImprove: Adapting Program Data to Deep Learning Models Through Program Transformations}
%%%\input{tex/story}
%Codeimprove: Program Adaptation for Deep Code Models
%\input{tex/abstract}
\maketitle
\input{tex/abstract}
\input{tex/introduction}

\input{tex/research_approach}

\input{tex/contributions}

\input{tex/preliminary_results}

%\input{tex/background}

%\input{tex/preliminary}
%\input{tex/design}

%\input{tex/evaluation}
%\input{tex/limitation}

%\input{tex/threats}

%\input{tex/applications}

%\input{tex/related}

%\input{tex/conclusion}

%\input{tex/dataavailability}

%% Acknowledgments

\clearpage
\bibliographystyle{IEEEtran}
\bibliography{conference_101719}
\end{document}

%% file: tex/macros.tex
\newcommand{\todo}[1]{\textbf{\color{red}{Ravishka: #1} }}

\newcommand{\wei}[1]{\textcolor{blue}{Wei: [#1]}}

\newcommand{\ww}[1]{\textcolor{blue}{Weihang: [#1]}}

\newcommand{\td}[1]{\textcolor{blue}{ToDo: [#1]}}

\newcommand{\zijie}[1]{{\color{purple}[Zijie: #1]}}

\newcommand{\zeqing}[1]{{\textcolor{blue}[Zeqing: #1]}}

\newcommand{\jiajun}[1]{{\color{cyan}[Jiang: #1]}}

\newcommand{\distance}{4pt}
\setlength{\textfloatsep}{\distance}

\newcommand\mycommfont[1]{\small\ttfamily\textcolor{violet}{#1}}
\SetCommentSty{mycommfont}

\lstdefinestyle{Cpp}{ % Define a style for your code snippet, multiple definitions can be made if, for example, you wish to insert multiple code snippets using different programming languages into one document
	%    backgroundcolor=\color{highlight}, % Set the background color for the snippet - useful for highlighting
	language=C++,
	basicstyle=\scriptsize\ttfamily, % The default font size and style of the code
	breakatwhitespace=false, % If true, only allows line breaks at white space
	breaklines=true, % Automatic line breaking (prevents code from protruding outside the box)
	captionpos=b, % Sets the caption position: b for bottom; t for top
	commentstyle=\color[rgb]{0.0, 0.5, 0.69},%\color[rgb]{0,0.6,0}, % Style of comments within the code - dark green courier font
	deletekeywords={}, % If you want to delete any keywords from the current language separate them by commas
	%escapeinside={\%}, % This allows you to escape to LaTeX using the character in the bracket
	escapeinside={<@}{@>},
	firstnumber=1, % Line numbers begin at line 1
	frame=lines, % Frame around the code box, value can be: none, leftline, topline, bottomline, lines, single, shadowbox
	frameround=tttt, % Rounds the corners of the frame for the top left, top right, bottom left and bottom right positions
	keywordstyle={[1]\color{blue!90!black}},
	keywordstyle={[3]\color{red!80!orange}},
	morekeywords={String,int}, % Add any functions no included by default here separated by commas
	numbers=left, % Location of line numbers, can take the values of: none, left, right
	numbersep=-8pt, % Distance of line numbers from the code box
	numberstyle=\tiny\color[rgb]{0.1,0.1,0.1}, % Style used for line numbers
	rulecolor=\color{black}, % Frame border color
	showstringspaces=false, % Don't put marks in string spaces
	showtabs=false, % Display tabs in the code as lines
	stepnumber=1, % The step distance between line numbers, i.e. how often will lines be numbered
	stringstyle=\color[rgb]{0.58,0,0.82},
	tabsize=2, % Number of spaces per tab in the code
	backgroundcolor=\color{white}
}

%% file: tex/abstract.tex
\begin{abstract}
%\textcolor{red}{}

Advancements in deep learning have significantly improved model performance across tasks involving code, text, and image processing. However, these models still exhibit notable mispredictions in real-world applications, even when trained on up-to-date data. Such failures often arise from slight variations in inputs such as minor syntax changes in code, rephrasing in text, or subtle lighting shifts in images that reveal inherent limitations in these models’ capability to generalize effectively. Traditional approaches to address these challenges involve retraining, a resource-intensive process that demands significant investments in data labeling, model updates, and redeployment.

This research introduces an adaptive, on-the-fly input refinement framework aimed at improving model performance through input validation and transformation. The input validation component detects inputs likely to cause errors, while input transformation applies domain-specific adjustments to better align these inputs with the model’s handling capabilities. This dual strategy reduces mispredictions across various domains, boosting model performance without necessitating retraining. As a scalable and resource-efficient solution, this framework holds significant promise for high-stakes applications in software engineering, natural language processing, and computer vision.

\end{abstract}

%% file: tex/introduction.tex
\section{Introduction}

Deep learning models have achieved remarkable success across domains such as software engineering, natural language processing (NLP), and computer vision~\cite{tian2023fly, naturalattack, Zhang2023Challenging, lu2021codexglue,svyatkovskiy2020intellicode,Devlin2019BERT,Sanh2020DistilBERT,guo2021graphcodebert,Chen2021Evaluating,xiao2022repairing,xiao2021selfchecking}. Generally, these models are trained on large, current datasets and evaluated on test data that closely mirrors the training distribution. However, despite this alignment, models frequently mispredict in real-world applications, not only due to distribution shifts but also due to their inability to manage slight input variations within the same distribution.

%%Deep learning models have achieved remarkable success across diverse domains such as software engineering, natural language processing (NLP), and computer vision. These models are typically trained on large, up-to-date datasets and evaluated on test data that aligns closely with the training distribution. However, despite this alignment, models frequently mispredict in real-world applications. These failures occur not only due to shifts in data distribution but because models struggle to handle nuanced, subtle variations present even within the same data distribution. %This phenomenon highlights a fundamental limitation in deep learning: \textit{models learn patterns based on statistical correlations, yet they lack the contextual understanding and adaptability necessary to reliably interpret small but meaningful differences in input data}.

A fundamental limitation of deep learning models is their reliance on statistical correlations rather than true understanding, restricting their ability to consistently interpret even familiar inputs. For instance, code language models (CLMs) often mispredict when confronted with minor syntactic changes, such as variable renaming or slight control flow adjustments~\cite{tian2023fly, naturalattack, Zhang2023Challenging}. These models learn code as sequential patterns, lacking an understanding of underlying functionality, making them fragile to variations that are functionally equivalent.

%Deep learning models are fundamentally constrained by their dependence on statistical correlations rather than intrinsic understanding, limiting their ability to consistently interpret inputs—even those aligned with training data distributions. In code language models (CLMs), for instance, slight syntactic modifications, such as variable renaming or marginal control flow adjustments, can lead to mispredictions because these models lack a true grasp of underlying functionality and dependencies. They perceive code primarily as a sequence of patterns rather than a cohesive logical structure, resulting in fragile predictions sensitive to minor, functionally equivalent variations.

Similarly, NLP models capture linguistic patterns but lack human-like adaptive interpretation of context, tone, or intent~\cite{chowdhary2020natural}. Slight changes in phrasing or word emphasis, even within familiar contexts, can yield different meanings that these models fail to understand especially in nuanced tasks like sentiment analysis and dialogue systems. Image-based models~\cite{xiao2022repairing,xiao2021selfchecking, wang2020dissector} are equally vulnerable, as slight perturbations in lighting or angle can lead to mispredictions due to their reliance on precise feature mappings rather than broader perceptual generalization.

%%In natural language processing (NLP), models capture linguistic patterns but lack the adaptive interpretive capacity that human cognition applies to context, tone, and intent. Even when inputs resemble those seen in training, subtle shifts in phrasing or word emphasis can produce divergent meanings that these models fail to apprehend, particularly in nuanced applications like sentiment analysis and dialogue systems.

%%Similarly, image-based models are highly susceptible to minor perturbations in visual inputs, such as lighting shifts or angle adjustments, due to their dependence on precise feature mappings. Although trained on diverse visual data, these models often misclassify inputs that deviate slightly from learned representations, as they cannot generalize beyond narrowly defined perceptual patterns. 

These domain-specific limitations reveal a broader challenge: \textit{while deep learning models excel at reproducing complex statistical relationships, they lack the flexibility, contextual reasoning, and adaptability needed to handle the complexities inherent in real-world data}. \textbf{Retraining}~\cite{yuDataAugmentationProgram2022,xiao2021selfchecking,xiao2022repairing}, a traditional approach to addressing these issues which involves periodic model updates to account for new or rare patterns. However, retraining is resource-intensive, requiring substantial computational power, time, and labeled data. Additionally, retraining can lead to overfitting on new data and is, at best, a temporary solution for handling the infinite variations of real-world inputs. These limitations underscore the need for a more adaptive approach in deep learning systems.

To address these challenges, this research introduces an adaptive framework for \textbf{input refinement}. Rather than relying on retraining, this framework improves model performance by dynamically performing \textbf{input validation} and \textbf{input transformation} during inference. Through a structured sequence of phases, this approach provides a scalable and resource-efficient solution for reducing mispredictions without modifying the model itself.

%%To overcome the limitations of traditional deep learning approaches in handling  input variability, this research presents a comprehensive framework for \textbf{input refinement}. This framework directly addresses the challenge of misprediction by equipping models with the capability to \textbf{input validation} and \textbf{input transformation} dynamically at inference. Through a structured sequence of phases, the framework provides a scalable, resource-efficient solution for enhancing model reliability without relying on retraining.

The framework consists of three main phases, each targeting specific challenges. The first phase, \textbf{Input Validation (P1)}, identifies inputs likely to cause errors, even if they are within the model’s training distribution. Tailored to various architectures, this phase includes sub-phases for validation encoder models \textbf{(P1.1)}, validation for decoder models \textbf{(P1.2)}, and validation for encoder-decoder models \textbf{(P1.3)}.

%%The framework consists of three main phases, each with specialized sub-phases designed to address specific challenges. \textbf{Input Validation (P1)} focuses on identifying inputs that are likely to produce wrong predictions, even if they technically fall within the model’s data distribution. This phase examines the unique architectural behavior of inputs in various model types, ensuring validation is tailored to the specific processing stages of each architecture. Input validation includes three sub-phases: validation for encoder models \textbf{(P1.1)}, validation for decoder models\textbf{(P1.2)}, and validation for encoder-decoder models \textbf{(P1.3)}.
%By evaluating layer-wise consistency, each sub-phase identifies subtle input variations that may destabilize the model’s interpretive accuracy.

The second phase, \textbf{Input Transformation (P2)}, adjusts inputs identified "out-of-scope" in the validation phase, applying domain-specific modifications to ensure alignment with the model's handling capabilities. This phase includes discrete transformations for code \textbf{(P2.1)} and text \textbf{(P2.2)}, as well as continuous transformations for images \textbf{(P2.3)}, preserving the intrinsic meaning or functionality of inputs while ensuring they align more closely with the model’s learned patterns.

%%\textbf{Input Transformation (P2)} targets the out-of-scope inputs from the validation phase, applying domain-specific transformations to adapt them at the data level. This phase addresses the unique properties of different data types through two primary sub-phases: discrete data transformation (for code \textbf{(P2.1)} and text \textbf{(P2.2)}) and continuous data refinement (for images \textbf{(P2.3)}). These refinements adjust inputs in a way that preserves their intrinsic meaning or functionality, ensuring they align more closely with the model’s learned patterns and processing strengths.

The final phase, \textbf{Optimal Search Strategy (P3)}, iterates over transformed input variations to find the best representation for model processing. This adaptive search further reduces mispredictions without changing model parameters or requiring retraining.

%%The final phase, \textbf{Optimal Search Strategy (P3)}, further enhances adaptability by iterating over refined versions of out-of-scope inputs to determine the best representation for model processing. This phase involves a search sub-phase that dynamically selects the optimal transformation, thereby reducing mispredictions without altering model parameters or requiring retraining.

This research hypothesizes that a framework incorporating input validation, targeted transformation, and optimal search will improve deep learning models' performance across code, NLP, and image domains. By dynamically aligning inputs with learned model patterns, this approach aims to significantly reduce mispredictions in real-world settings without requiring retraining, offering a scalable and efficient solution for robust AI performance in complex environments.

\begin{figure*}[!htbp]
\centering
\includegraphics[width=\linewidth]{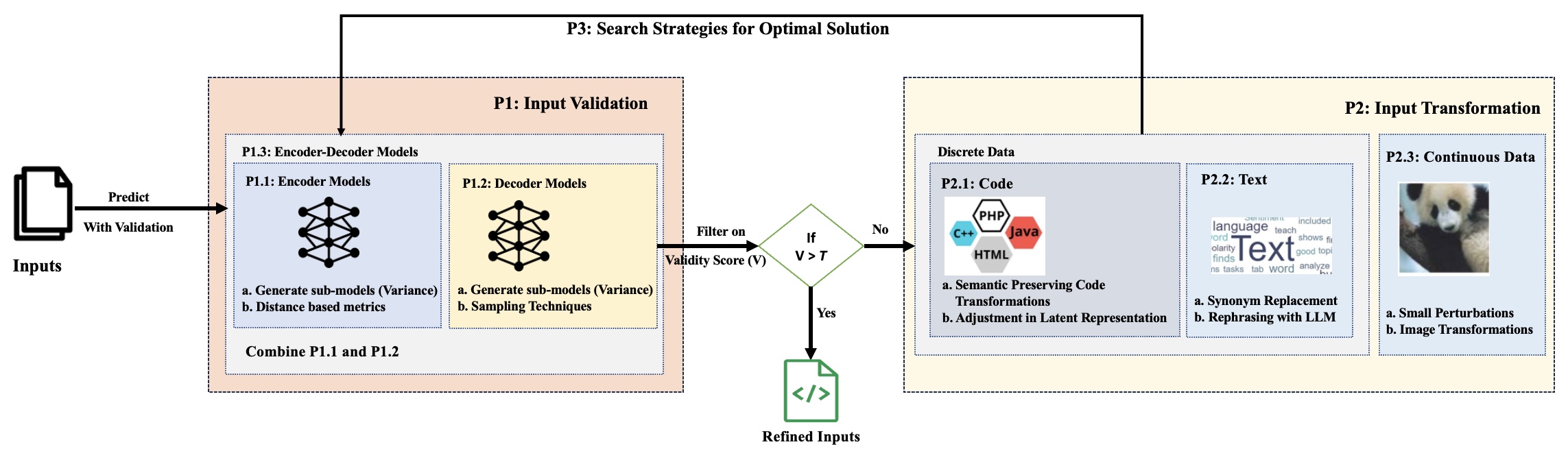}
\caption{Overview of Proposed On the Fly Input Refinement Framework} 
% %\textcolor{red}{TODO: DRAW A FLOW DIAGRAM TO SHOW MANIFESTATION }
\label{fig:Overview}
\end{figure*}

%% file: tex/research_approach.tex
\section{Research Overview}

Figure~\ref{fig:Overview} provides an overview of the proposed framework, which addresses the ongoing challenge of mispredictions in deep learning models across code, natural language processing (NLP), and image-based tasks. This framework introduces three key phases: \textbf{input validation (P1), input transformation (P2)}, and \textbf{optimal search strategy (P3)} to improve model performance during inference, thereby reducing dependency on costly retraining. Each phase employs distinct mechanisms tailored to different data domains, making the approach both adaptable and comprehensive.

%%%Figure~\ref{fig:Overview}  provides an overview of my proposed research. This research proposes a novel framework to address the persistent challenge of mispredictions in deep learning models, particularly in the nuanced environments of code, natural language processing (NLP), and image-based tasks. The framework leverages input validation \textbf{(P1)}, transformation \textbf{(P2)}, and an adaptive search strategy \textbf{(P3)} to improve the model performance at inference, reducing dependency on retraining. Each phase contributes a distinct mechanism to improve model reliability across different data domains, making this approach both comprehensive and flexible.

\textbf{P1: Input Validation.} The objective of \textbf{(P1)} is to identify inputs likely to produce unreliable predictions (out-of-scope inputs), even if they fall within the model's training distribution. To accommodate the unique characteristics of various model architectures, this phase is organized into three sub-phases: validations for encoder models \textbf{(P1.1)}, decoder models \textbf{(P1.2)}, and encoder-decoder models \textbf{(P1.3)}.

%\textbf{P1: Input Validation.} The goal of \textbf{(P1)} is to identify inputs that may lead to unreliable predictions (i.e., out-of-scope), even if they align with the model’s learned data distribution. This phase is designed to address the unique characteristics of various model architectures, employing specialized techniques for each type. Input Validation is structured into three sub-phases:  approaches for encoder models \textbf{(P1.1)}, decoder models \textbf{(P1.2)}, and encoder-decoder models \textbf{(P1.3)}.

 The \textbf{P1.1} sub-phase applies a set of sub-models to assess input consistency through variance and distance metrics within data representations. By generating slightly varied sub-models, this approach evaluates how an input propagates across layers, examining consistency in token representation and feature extraction. In addition, a distance-based metric is employed to assess the coherence of token propagation, identifying anomalies that may signal out-of-scope inputs. Through these variance and distance-based metrics, encoder models gain an enhanced capacity to detect inputs that introduce discrepancies in hidden state representations, thereby improving validation.
 
 %%For \textbf{P1.1}, input validation utilizes a set of sub-models that analyze variance and distance metrics within the data representation. By generating sub-models with slight variations, this approach can examine how consistently an input propagates through each layer, detecting potential discrepancies in token representation or syntactic structure. Additionally, a distance-based approach can be applied to assess the coherence of token propagation, identifying anomalies in the encoded representation that might signal an out-of-scope input. This variance and distance-based validation approach allows encoder models to flag inputs that introduce inconsistencies in foundational representations. 

 The \textbf{P1.2} sub-phase where the primary task involves generating sequences or predictions, the validation process focuses on identifying inconsistencies in output generation. This sub-phase uses sampling techniques to evaluate how each generated sequence or prediction changes across layers. By measuring the variance of generated outputs and detecting unexpected deviations, this method flags inputs that are prone to lead to unstable or incorrect generations. Sampling-based validation allows for early detection of inconsistencies in the decoding process, ensuring each generated output remains stable and aligned with intended meanings.

 % In \textbf{P1.2}, where the primary task often involves generating sequences or predictions, validation is focused on identifying inconsistencies in output generation. To accomplish this, validation checks how each generated sequence or prediction varies across layers, employing sampling techniques to evaluate consistency at each stage. By measuring the variance of generated outputs and identifying unexpected deviations, this sub-phase can highlight inputs that are likely to lead to unstable or incorrect generation. Sampling-based approaches allow the model to detect generation inconsistencies early in the decoding process, ensuring that each generated output remains stable and aligned with the intended meaning. 

The \textbf{P1.3} sub-phase combines elements from both \textbf{P1.1} and \textbf{P1.2} to provide a comprehensive validation framework. This integration enables consistency checks across both encoding and decoding stages, tracking how encoded representations are transformed during decoding. By merging validation techniques from encoder and decoder models, this approach effectively detects inputs that could cause conflicts between encoding and decoding interpretations, thereby improving the robustness of validation in complex architectures.

%%%For \textbf{P1.3}, validation integrates checks across both \textbf{P1.2} and \textbf{P1.2} to provide a comprehensive assessment of each input’s validation. This sub-phase incorporates consistency evaluations in both the encoding and decoding layers, tracking how encoded representations are transformed during decoding. By combining the strengths of both \textbf{P1.1} and \textbf{P1.2} validation techniques, this approach improves the model’s ability to detect inputs that introduce conflicts between encoding and decoding interpretations, resulting in more robust validation for complex architectures.

\textbf{P2: Input Transformation.}  The \textbf{P2} addresses inputs flagged as out-of-scope during validation by applying targeted transformations to align these inputs more closely with the model’s handling capabilities. This phase is divided into two main categories based on data type: \textbf{Discrete Data Transformation (Code (P2.1)} and \textbf{Text (P2.2)}) and \textbf{Continuous Data Transformation (P2.3)}. Each category is designed to preserve the input’s original meaning or function while making adjustments to reduce interpretive inconsistencies. Tailoring transformations to each data type ensures that inputs are modified in ways that manage complexity without altering essential content.

%addresses inputs that are identified as out-of-scope  during validation by applying targeted transformations to better align these inputs with the model’s handling capability. This phase is divided into two main categories based on data type: Discrete Data Transformation (\textbf{P2.1} and \textbf{P2.2}) and Continuous Data Transformation (\textbf{P2.3}). Each category is designed to preserve the input’s original meaning or function while adjusting it to reduce interpretive inconsistencies. By tailoring transformations to each data type, this phase ensures that inputs are transformed in ways that mitigate their complexity without altering their essential content.

Discrete Data Transformation includes structured data like code and text, where maintaining original meaning and functionality is crucial. This transformation uses specific techniques to make minor adjustments that reduce interpretive challenges without changing the intended purpose of the data.

%%Discrete data includes structured information like code and text, where maintaining the original meaning or functionality is critical. For both data types, transformation techniques ensure that minor modifications reduce interpretive challenges without altering the data’s intended purpose.

For \textbf{P2.1} sub-phase focuses on semantic-equivalent transformations and latent space adjustments. Semantic-equivalent transformations modify code syntax or structure while preserving its functionality, helping to clarify complex or ambiguous constructs. For instance, a conditional loop could be refactored into a functional equivalent, or redundant sections could be simplified for improved clarity. Additionally, latent space transformations adjust the encoded representation of code, enabling the model to interpret structural nuances more effectively. Together, these transformations ensure that code inputs align with the model’s learned patterns, reducing the risk of misinterpretation due to syntactic variability.

%For \textbf{P2.1} refinement focuses on semantic-equivalent transformations and adjustments in the latent space. Semantic-equivalent transformations apply modifications that preserve the code’s functionality but alter its syntax or structure, which can help clarify complex or ambiguous constructs. For example, a conditional loop could be refactored into a functional equivalent, or redundant code could be simplified to improve clarity. Additionally, latent space transformations adjust the encoded representation of code, enabling the model to better interpret structural nuances. These transformations ensure that code inputs remain aligned with the model’s learned patterns, minimizing the chance of misinterpretation due to syntactic variability.

In \textbf{P2.2}  sub-phase applies techniques such as synonym replacement and text rephrasing with large language models (LLMs). Synonym replacement changes specific words within the text to reduce ambiguity or simplify language without altering the overall meaning. For instance, in a sentiment analysis task, “joyful” might replace “happy” to match the model’s vocabulary. Additionally, LLM-based rephrasing enables comprehensive sentence or phrase rewording to standardize informal or complex language. For example, “I’m over the moon about this product” could be rephrased to “I’m extremely happy with this product,” making it more compatible with the model’s expected input patterns. To ensure transformations align with the model’s learned vocabulary, we use token probability distributions and reranking based on log likelihood, refining selections beyond greedy or beam search. These text transformations help reduce interpretive errors by bringing language closer to the model’s trained structure.

%In \textbf{P2.2}, transformation uses techniques such as synonym replacement and text rephrasing with large language models (LLMs). Synonym replacement adjusts specific words within the text to reduce ambiguities or simplify language, without changing the overall meaning. For example, in a sentiment analysis task, the word "joyful" might replace "elated" to match the model’s training vocabulary. Additionally, leveraging LLMs enables full rephrasing of sentences or phrases to standardize informal or complex language. For instance, "I’m over the moon about this product" could be rephrased to "I’m extremely happy with this product," aligning with more formal patterns that the model may interpret more reliably. These text refinements help mitigate interpretive errors by bringing language closer to the model’s expected input structure. 

Continuous Data Transformation targets high-dimensional data like images, which are sensitive to environmental factors.  In \textbf{P2.3} sub-phase includes small perturbations and transformations to normalize aspects such as lighting, contrast, and angle, stabilizing the model’s interpretation of visual features. This includes brightness adjustments can address counteract poor lighting, contrast normalization can make subtle details more noticeable, and geometric transformations like slight rotations or cropping to standardize perspective. For example, brightness adjustments can address lighting inconsistencies, while contrast normalization increases feature visibility. Geometric transformations ensure that objects remain within the model’s learned visual range, regardless of minor changes in viewpoint. Additionally, noise reduction techniques remove sensor artifacts, creating a cleaner input that reduces the likelihood of misclassification. These transformations establish a consistent visual baseline, allowing the model to interpret high-dimensional visual data more effectively.

%%%Continuous data, such as images, presents unique challenges due to its high-dimensional nature and sensitivity to environmental factors.  In \textbf{P2.3}, small perturbations or transformations help normalize lighting, contrast, and angle to stabilize the model’s interpretation of visual features. For example, brightness adjustments can counteract the effects of poor lighting, while contrast normalization makes subtle details more discernible. Geometric transformations like slight rotations or cropping help standardize perspective, ensuring that objects remain within the model’s learned visual range. Additionally, noise reduction techniques can remove sensor artifacts, creating a cleaner input that reduces the likelihood of misclassification. These transformations provide continuous data with a consistent visual baseline, allowing models to interpret high-dimensional inputs more effectively.

\textbf{P3: Search Strategies.} \textbf{P3} combines the insights from \textbf{P1} and \textbf{P2} to systematically identify the optimal modifications that convert out-of-scope inputs into forms the model can reliably interpret. This phase leverages different search-based approaches such as greedy, beam, and evolutionary searches to evaluate and select the best transformation for each input type, whether through syntax adjustments in code, rephrasing in text, or visual enhancements in images.

An adaptive search strategy is crucial for aligning complex inputs with the model’s learned patterns, as it allows for the dynamic exploration of transformation options. By iteratively refining inputs through this search process, the framework ensures reliable model performance on real-world data without the need for retraining.

%% file: tex/contributions.tex
\section{Contributions}

In summary, this research makes the following contributions: 

\begin{itemize}
    \item \textbf{Development of Multi-Level Validation Metrics}: This work introduces a comprehensive set of validation metrics to detect out-of-scope inputs across diverse model architectures, including encoder, decoder, and encoder-decoder models. These metrics support tasks in code, NLP, and image domains, enabling precise identification of inputs prone to misprediction.
    %My research introduces a set of validation metrics designed to detect out-of-scope inputs across different model architectures, including encoder models, decoder models, and encoder-decoder models. These metrics address diverse tasks across code, NLP, and image domains, enabling more precise identification of inputs likely to result in mispredictions.
    \item \textbf{Automated Input Transformation Techniques}: The proposed framework includes semantic-preserving input transformation techniques tailored to each data type (discrete and continuous), effectively aligning out-of-scope inputs with model capabilities.
    %My proposed framework includes effective input transformation techniques that apply semantic-preserving transformations tailored to each data type (discrete and continous). 
    \item \textbf{Adaptive Search Strategy}: An innovative search strategy dynamically identifies the optimal refinements for out-of-scope inputs, converting them into in-scope representations and improving model performance without retraining.
    %My research introduces a novel search strategy to dynamically identify the optimal refinement for out-of-scope inputs, allowing the model to convert out-of-scope inputs into in-scope representations. 
    \item \textbf{Comprehensive Evaluation Framework}: Extensive testing across various model architectures and tasks demonstrates the framework’s scalability and adaptability in real-world applications.
    %Extensive testing is conducted across various model architectures and tasks, demonstrating the scalability and adaptability of the framework in real-world applications.
    \item \textbf{Scalable, Resource-Efficient Solution for Model Deployment}: By dynamically refining inputs at inference, this framework offers a scalable, resource-efficient alternative to frequent retraining, making it well-suited for agile development and large-scale deployment, while reducing operational costs and maintaining model performance.
    %By providing a framework that dynamically transforms inputs at inference, this research offers a scalable and resource-efficient alternative to frequent retraining. This solution is ideal for agile development environments and large-scale deployment scenarios, reducing costs and improving model performance on an ongoing basis.
    \item \textbf{Publications and Open Science Contributions}: This research advances the field through publications in leading AI and software engineering venues. Findings and public artifacts will be shared in alignment with open science principles, supporting reproducibility, transparency, and adherence to high research standards.
    %This work contributes to the field through publications in leading AI and software engineering venues. Research findings, along with public artifacts, are shared in alignment with open science principles, ensuring reproducibility, transparency, and adherence to high research standards.
\end{itemize}

%% file: tex/preliminary_results.tex
\section{Research Progress}

Our preliminary framework addresses key components of the input refinement process, specifically in \textbf{P1.1} (Input Validation for Code-based Encoder Models), \textbf{P2.1} (Input Transformation for Code Data), and \textbf{P3} (Optimal Search via Adaptation by Evolutionary Search). Results from this framework have been accepted at ICSE 2025~\cite{rathnasuriya2025codeimprove}.

%Our preliminary framework addresses key components of the input refinement process by solving for \textbf{P1.1} (Input Validation for Code-based Encoder Models), \textbf{P2.1} (Input Transformation for Code Data), and \textbf{P3} (Optimal Search via Adaptation by Evolutionary Search). The results of this framework have been accepted at ICSE’25. 

\textbf{P1.1} introduces a novel layerwise validation approach tailored for encoder architectures, addressing limitations of traditional uncertainty metrics~\cite{guo2017calibration,wang2020dissector,hendrycks2018baseline,gal2016dropout, alon2019code2vec,xiao2019quantifying,vasudevan2019towards,corbiere2019addressing, monarch2021human, steinhardt2016unsupervised,shannon1948mathematical} with structured, discrete data like code. We developed a Dropout-Based Sub-model Generation (DSMG) technique to capture layerwise processing of nuanced structural elements that includes variable dependencies and control flow. This enhances uncertainty estimation, enabling more accurate classification of out-of-scope inputs.

%\textbf{P1.1} focuses on identifying out-of-scope inputs by introducing a novel, layer-sensitive validation approach specific to encoder architectures. Traditional uncertainty metrics often fall short with structured, discrete data like code. To address this, our method employs a set of sub-models that capture layer-wise dynamics across the encoder, isolating inconsistencies in the representation of out-of-scope inputs. We developed a novel, Dropout-Based Sub-model Generation (DSMG) technique to examine layer-specific processing, this approach detects nuanced structural elements such as variable dependencies and control flow, improving uncertainty estimation and enabling more accurate classification of input scope. 

\textbf{P2.1} applies semantic-preserving transformations to align out-of-scope code inputs with the model’s learned representations without altering functional intent. Transformations include restructuring and reordering statements to maintain operational meaning while improving compatibility with the model’s processing. Our framework includes 15 semantic-preserving code transformation operators. 

%\textbf{P2.1} applies semantic-preserving transformations to out-of-scope inputs, aligning them with the model’s learned representations without altering functional intent. For code data, these transformations include restructuring or reordering code statements in ways that maintain their operational meaning, enhancing compatibility with the model’s internal processing. 

\textbf{P3} utilizes an Adaptation by Evolutionary Search (AES) strategy to iteratively identify the best possible refinement for each out-of-scope inputs by applying a sequence of semantic-preserving transformations and sampling adjustments guided by DSMG validity scores. AES enables the dynamic evolution of out-of-scope inputs into compatible forms at inference time, thereby improving the model's performance without retraining.

%For \textbf{P3}, my work utilizes an Adaptation by Evolutionary Search (AES) strategy to iteratively identify the best possible refinement for each out-of-scope input. Guided by DSMG-based validity scores, AES applies a sequence of semantic transformations and sampling adjustments, evolving flagged inputs into in-scope forms that the model can reliably process. This adaptive search strategy allows for dynamic refinement at inference, ensuring robust model performance without the need for retraining.

The experimental evaluation of this proposed framework is conducted on encoder models, specifically CodeBERT~\cite{fengCodeBERTPreTrainedModel2020}, RoBERTa~\cite{Liu2019RoBERTa}, and GraphCodeBERT~\cite{guo2021graphcodebert} . The research assess  effectiveness across various software engineering tasks, focusing on classification tasks (vulnerability detection and defect prediction). For each task, the research use well-established datasets such as devign dataset ~\cite{zhou2019devign} for vulnerability detection and CodeChef dataset~\cite{phan2017conv} for defect prediction. 

%My approach is evaluated based on specific research questions (RQs) to examine its impact on model performance, out-of-scope data detection, input refinement effectiveness (only using semantic preserving code transformations), and overall efficiency. The results of this framework have been accepted at ICSE’25.

\subsection{Overall Performance}
This proposed work is compared with Input-Reflector~\cite{xiao2022repairing}, a similar technique to used to repair failure inducing inputs in image data. I adapted Input-Reflector to code domain. Notably, the findings include: (1) This work achieved the best model improvement ranging upto 8.78\% in accuracy, 8.48\% in precision,
16.9\% in recall, and 13.5\% in F1-score on all the subjects; (2) My work is successfully capable of correcting around 23.1\% to 39.9\% of the mispredicted inputs on both vulnerability detection and defect prediction tasks; (3) Input-Reflector cannot be applied to code data. Our results indicate that Input-Reflector negatively impacts the model performance for the CodeBERT model on both vulnerability detection and defect prediction tasks; (4) The negative impact of this framework is minimal. At most, this framework will only mispredict 2.6\% of the correct predictions to become mispredictions. 

\subsection{Effectiveness of Input Validation}
\label{validation_results}

This frameworks’s DSMG is compared with the
Cross-Layer Dissection (CLD)~\cite{wang2020dissector}. Additionally, I evaluated the DSMG approach with the uncertainty metrics: vanilla~\cite{hendrycks2018baseline}, temperature scaling~\cite{guo2017calibration}, predictive entropy~\cite{steinhardt2016unsupervised,shannon1948mathematical}, entropy~\cite{steinhardt2016unsupervised,shannon1948mathematical}, mutual information~\cite{steinhardt2016unsupervised,shannon1948mathematical}, least confidence~\cite{monarch2021human}, ratio confidence~\cite{monarch2021human}, and margin confidence~\cite{monarch2021human}, monte-carlo dropout~\cite{gal2016dropout}, and deep ensemble~\cite{lakshminarayanan2017simple}. 

Based on the results, the findings include: (1) DSMG achieved higher AUC scores across all models and tasks (i.e., AUC 0.781- 0.924) compared to uncertainty metrics which achieved an AUC score of 0.624; (2) The Correction validation rate (CVR), i.e., the successfully validated misprediction on DSMG is higher than all other baselines. Moreover, DSMG can detect 70.4\% of the out-of-scope inputs for the CodeBERT model on vulnerability detection tasks; (3) The Mis-correction validation rate (MVR), i.e., correct predictions validated
as mispredictions on DSMG is lower than other approaches, concluding that DSMG is better
at differentiating in-scope inputs. MVR for defect prediction task shows 3.0\%, 3.1\%, and 3.3\% for CodeBERT, RoBERTa, and GraphCodeBERT models; (4) other uncertainty metrics did not produce promising results on AUC, CVR, or MVR. For example, predictive entropy obtained a CVR of 43.5\% and an MVR of 35.3\%, which are not significant indicators of effective performance.

\subsection{Effectiveness of Search Strategies}
The proposed AES algorithm is compared with  two search strategies; namely random search (Rand)~\cite{zabinsky2009random} that applies random transformations until identifying the optimal candidate and Hill climbing (HC) algorithm~\cite{selman2006hill}. 

AES algorithm achieved the highest accuracy across all evaluated tasks, with improvements of up to 8.78\%, outperforming both Rand and HC, which showed gains of only up to 2.13\%. While Rand and HC improve model performance, their search algorithms are limited by early stopping at local minima. These algorithms terminate as soon as they identify an improved candidate, whereas AES continues evolving candidates over multiple generations to achieve the best possible outcome. 

\subsection{Semantic Integrety and Overhead}
The proposed framework’s refinement techniques are specifically designed to maintain the semantic integrity of code inputs. In evaluations, the transformations consistently preserve original code functionality, a key advantage over alternative methods, which may alter functionality. 

The proposed framework also achieves competitive runtime efficiency, with transformation rates (TPS) between 1.2 and 2.04 TPS. Online adaptation times per input range from 49.92 to 59.4 seconds, demonstrating manageable overheads for practical deployment. Moreover, this framework is more efficient and offers a practical, scalable solution to enhance model performance without the significant cost and time investment required by traditional methods such as retraining and model replacement.

\section{Timeline for completion}
I plan to complete the \textbf{P1.2, P2.3} along with \textbf{P3} of this project—refining inputs specifically for generative tasks—by January 2025. Following this, I will incorporate \textbf{P2.2}, with an expected completion by February 2025. By the end of March 2025, my goal is to complete \textbf{P1.3}.

For the current framework, I will focus on expanding transformation rules, reducing transformation times, and refining the validation metric, with the goal of finalizing the complete framework by March 2025. My primary aim is to submit several papers to top-tier conferences throughout this period. Once these project phases are completed, I plan to finish writing my dissertation and aim to defend my doctoral thesis by December 2025.

%I plan to complete the second phase of this project on refining inputs for generative tasks by January 2025. This phase incudes adopting the DSMG validation metric towards decoder-only models and incorporating sampling techniques to input refinement search. Meanwhile I will incorporate both encoder-only model's technique and decoder-only model's technique to input refinement for encoder-decoder only models. This step is expected to complete by February 2025. Also, for my current approach, I will work on adding more transformation rules, minimizing transformation times, and improve the validation metric. All of this framework is planned for complete by March 2025. The goal is to submit several Top Tier conference publications. Once I complete these tasks, my plan is to finish writing the dissertation and defend my doctoral thesis by December 2025.

%% file: conference_101719.bbl
\begin{thebibliography}{10}
\providecommand{\url}[1]{#1}
\csname url@samestyle\endcsname
\providecommand{\newblock}{\relax}
\providecommand{\bibinfo}[2]{#2}
\providecommand{\BIBentrySTDinterwordspacing}{\spaceskip=0pt\relax}
\providecommand{\BIBentryALTinterwordstretchfactor}{4}
\providecommand{\BIBentryALTinterwordspacing}{\spaceskip=\fontdimen2\font plus
\BIBentryALTinterwordstretchfactor\fontdimen3\font minus \fontdimen4\font\relax}
\providecommand{\BIBforeignlanguage}[2]{{%
\expandafter\ifx\csname l@#1\endcsname\relax
\typeout{** WARNING: IEEEtran.bst: No hyphenation pattern has been}%
\typeout{** loaded for the language `#1'. Using the pattern for}%
\typeout{** the default language instead.}%
\else
\language=\csname l@#1\endcsname
\fi
#2}}
\providecommand{\BIBdecl}{\relax}
\BIBdecl

\bibitem{tian2023fly}
Z.~Tian, J.~Chen, and X.~Zhang, ``On-the-fly improving performance of deep code models via input denoising,'' \emph{arXiv preprint arXiv:2308.09969}, 2023.

\bibitem{naturalattack}
\BIBentryALTinterwordspacing
Z.~Yang, J.~Shi, J.~He, and D.~Lo, ``Natural attack for pre-trained models of code,'' in \emph{Proceedings of the 44th International Conference on Software Engineering}, ser. ICSE '22.\hskip 1em plus 0.5em minus 0.4em\relax New York, NY, USA: Association for Computing Machinery, 2022, p. 1482–1493. [Online]. Available: \url{https://doi.org/10.1145/3510003.3510146}
\BIBentrySTDinterwordspacing

\bibitem{Zhang2023Challenging}
W.~Zhang, S.~Guo, H.~Zhang, Y.~Sui, Y.~Xue, and Y.~Xu, ``Challenging {{Machine Learning-based Clone Detectors}} via {{Semantic-preserving Code Transformations}},'' \emph{IEEE Transactions on Software Engineering}, vol.~49, no.~5, pp. 3052--3070, May 2023.

\bibitem{lu2021codexglue}
S.~Lu, D.~Guo, S.~Ren, J.~Huang, A.~Svyatkovskiy, A.~Blanco, C.~Clement, D.~Drain, D.~Jiang, D.~Tang \emph{et~al.}, ``Codexglue: A machine learning benchmark dataset for code understanding and generation,'' \emph{arXiv preprint arXiv:2102.04664}, 2021.

\bibitem{svyatkovskiy2020intellicode}
A.~Svyatkovskiy, S.~K. Deng, S.~Fu, and N.~Sundaresan, ``Intellicode compose: Code generation using transformer,'' in \emph{Proceedings of the 28th ACM joint meeting on European software engineering conference and symposium on the foundations of software engineering}, 2020, pp. 1433--1443.

\bibitem{Devlin2019BERT}
J.~Devlin, M.-W. Chang, K.~Lee, and K.~Toutanova, ``{{BERT}}: {{Pre-training}} of {{Deep Bidirectional Transformers}} for {{Language Understanding}},'' May 2019.

\bibitem{Sanh2020DistilBERT}
V.~Sanh, L.~Debut, J.~Chaumond, and T.~Wolf, ``{{DistilBERT}}, a distilled version of {{BERT}}: Smaller, faster, cheaper and lighter,'' Feb. 2020.

\bibitem{guo2021graphcodebert}
D.~Guo, S.~Ren, S.~Lu, Z.~Feng, D.~Tang, S.~Liu, L.~Zhou, N.~Duan, A.~Svyatkovskiy, S.~Fu, M.~Tufano, S.~K. Deng, C.~Clement, D.~Drain, N.~Sundaresan, J.~Yin, D.~Jiang, and M.~Zhou, ``Graphcodebert: Pre-training code representations with data flow,'' 2021.

\bibitem{Chen2021Evaluating}
M.~Chen, J.~Tworek, H.~Jun, Q.~Yuan, Pinto \emph{et~al.}, ``Evaluating {{Large Language Models Trained}} on {{Code}},'' Jul. 2021.

\bibitem{xiao2022repairing}
Y.~Xiao, Y.~Lin, I.~Beschastnikh, C.~Sun, D.~S. Rosenblum, and J.~S. Dong, ``Repairing failure-inducing inputs with input reflection,'' in \emph{The 37th IEEE/ACM International Conference on Automated Software Engineering (ASE)}.\hskip 1em plus 0.5em minus 0.4em\relax IEEE, 2022.

\bibitem{xiao2021selfchecking}
Y.~Xiao, I.~Beschastnikh, D.~S. Rosenblum, C.~Sun, S.~Elbaum, Y.~Lin, and J.~S. Dong, ``Self-checking deep neural networks in deployment,'' 2021.

\bibitem{chowdhary2020natural}
K.~Chowdhary and K.~Chowdhary, ``Natural language processing,'' \emph{Fundamentals of artificial intelligence}, pp. 603--649, 2020.

\bibitem{wang2020dissector}
H.~Wang, J.~Xu, C.~Xu, X.~Ma, and J.~Lu, ``Dissector: Input validation for deep learning applications by crossing-layer dissection,'' in \emph{Proceedings of the ACM/IEEE 42nd International Conference on Software Engineering}, 2020, pp. 727--738.

\bibitem{yuDataAugmentationProgram2022}
S.~Yu, T.~Wang, and J.~Wang, ``Data {{Augmentation}} by {{Program Transformation}},'' \emph{Journal of Systems and Software}, vol. 190, p. 111304, Aug. 2022.

\bibitem{rathnasuriya2025codeimprove}
R.~Rathnasuriya, Z.~Zhao, and W.~Yang, ``Codeimprove: Program adaptation for deep code,'' \emph{arXiv preprint arXiv:2501.15804}, 2025.

\bibitem{guo2017calibration}
C.~Guo, G.~Pleiss, Y.~Sun, and K.~Q. Weinberger, ``On calibration of modern neural networks,'' 2017.

\bibitem{hendrycks2018baseline}
D.~Hendrycks and K.~Gimpel, ``A baseline for detecting misclassified and out-of-distribution examples in neural networks,'' 2018.

\bibitem{gal2016dropout}
Y.~Gal and Z.~Ghahramani, ``Dropout as a bayesian approximation: Representing model uncertainty in deep learning,'' in \emph{international conference on machine learning}.\hskip 1em plus 0.5em minus 0.4em\relax PMLR, 2016, pp. 1050--1059.

\bibitem{alon2019code2vec}
U.~Alon, M.~Zilberstein, O.~Levy, and E.~Yahav, ``code2vec: Learning distributed representations of code,'' \emph{Proceedings of the ACM on Programming Languages}, vol.~3, no. POPL, pp. 1--29, 2019.

\bibitem{xiao2019quantifying}
Y.~Xiao and W.~Y. Wang, ``Quantifying uncertainties in natural language processing tasks,'' in \emph{Proceedings of the AAAI conference on artificial intelligence}, vol.~33, no.~01, 2019, pp. 7322--7329.

\bibitem{vasudevan2019towards}
V.~T. Vasudevan, A.~Sethy, and A.~R. Ghias, ``Towards better confidence estimation for neural models,'' in \emph{ICASSP 2019-2019 IEEE International Conference on Acoustics, Speech and Signal Processing (ICASSP)}.\hskip 1em plus 0.5em minus 0.4em\relax IEEE, 2019, pp. 7335--7339.

\bibitem{corbiere2019addressing}
C.~Corbi{\`e}re, N.~Thome, A.~Bar-Hen, M.~Cord, and P.~P{\'e}rez, ``Addressing failure prediction by learning model confidence,'' \emph{Advances in Neural Information Processing Systems}, vol.~32, 2019.

\bibitem{monarch2021human}
R.~M. Monarch, \emph{Human-in-the-Loop Machine Learning: Active learning and annotation for human-centered AI}.\hskip 1em plus 0.5em minus 0.4em\relax Simon and Schuster, 2021.

\bibitem{steinhardt2016unsupervised}
J.~Steinhardt and P.~S. Liang, ``Unsupervised risk estimation using only conditional independence structure,'' \emph{Advances in Neural Information Processing Systems}, vol.~29, 2016.

\bibitem{shannon1948mathematical}
C.~E. Shannon, ``A mathematical theory of communication,'' \emph{The Bell system technical journal}, vol.~27, no.~3, pp. 379--423, 1948.

\bibitem{fengCodeBERTPreTrainedModel2020}
Z.~Feng, D.~Guo, D.~Tang, N.~Duan, X.~Feng, M.~Gong, L.~Shou, B.~Qin, T.~Liu, D.~Jiang, and M.~Zhou, ``{{CodeBERT}}: {{A Pre-Trained Model}} for {{Programming}} and {{Natural Languages}},'' Sep. 2020.

\bibitem{Liu2019RoBERTa}
Y.~Liu, M.~Ott, N.~Goyal, J.~Du, M.~Joshi, D.~Chen, O.~Levy, M.~Lewis, L.~Zettlemoyer, and V.~Stoyanov, ``{{RoBERTa}}: {{A Robustly Optimized BERT Pretraining Approach}},'' Jul. 2019.

\bibitem{zhou2019devign}
Y.~Zhou, S.~Liu, J.~Siow, X.~Du, and Y.~Liu, ``Devign: Effective vulnerability identification by learning comprehensive program semantics via graph neural networks,'' \emph{Advances in neural information processing systems}, vol.~32, 2019.

\bibitem{phan2017conv}
A.~V. Phan and M.~Le~Nguyen, ``Convolutional neural networks on assembly code for predicting software defects,'' in \emph{2017 21st Asia Pacific Symposium on Intelligent and Evolutionary Systems (IES)}, 2017, pp. 37--42.

\bibitem{lakshminarayanan2017simple}
B.~Lakshminarayanan, A.~Pritzel, and C.~Blundell, ``Simple and scalable predictive uncertainty estimation using deep ensembles,'' \emph{Advances in neural information processing systems}, vol.~30, 2017.

\bibitem{zabinsky2009random}
Z.~B. Zabinsky \emph{et~al.}, ``Random search algorithms,'' \emph{Department of Industrial and Systems Engineering, University of Washington, USA}, 2009.

\bibitem{selman2006hill}
B.~Selman and C.~P. Gomes, ``Hill-climbing search,'' \emph{Encyclopedia of cognitive science}, vol.~81, p.~82, 2006.

\end{thebibliography}
